\documentclass[aps,amsmath,amssymb,amsfonts,superscriptaddress,nofootinbib]{revtex4}
\usepackage[english]{babel}

\newcommand{\bg}{{\bf g}}
\newcommand{\bk}{{\bf k}}
\newcommand{\bu}{{\bf u}}
\newcommand{\bv}{{\bf v}}
\newcommand{\br}{{\bf r}}

\newcommand\divergence{\operatorname{div}}

\newcommand{\beq}{\begin{equation}}
\newcommand{\eeq}{\end{equation}}
\newcommand{\bea}{\begin{eqnarray}}
\newcommand{\eea}{\end{eqnarray}}

\newcommand{\ga}{\mbox{${\gamma}$}}

\newcommand{\de}{\mbox{${\delta}$}}
\newcommand{\De}{\mbox{${\Delta}$}}

\newcommand{\si}{\mbox{${\sigma}$}}
\newcommand{\om}{\mbox{${\omega}$}}
\newcommand{\Om}{\mbox{${\Omega}$}}

\begin{document}
\selectlanguage{english}

%\title{Excitations created by small black hole passing through the Earth}
\title{Can one detect passage of small black hole through the Earth?}

\author{\firstname{I.~B.}~\surname{Khriplovich}}
\email{khriplovich@inp.nsk.su}
\author{\firstname{A.~A.}~\surname{Pomeransky}}
\email{a.a.pomeransky@inp.nsk.su}
 \affiliation{Budker Institute of
Nuclear Physics, 630090, Novosibirsk, Russia, and Novosibirsk
University}
\author{\firstname{N.}~\surname{Produit}}
\email{Nicolas.Produit@obs.unige.ch} \affiliation{INTEGRAL Science
Data Center, 16, Chemin d'Ecogia, CH-1290 Versoix, Switzerland}
\author{\firstname{G.~Yu.}~\surname{Ruban}}
\email{gennady.ru@gmail.com} \affiliation{Budker Institute of
Nuclear Physics, 630090, Novosibirsk, Russia, and Novosibirsk
University}

\begin{abstract}
The energy losses of a small black hole passing through the Earth
are examined. In particular, we investigate the excitations in the
frequency range accessible to modern acoustic detectors. The main
contribution to the effect is given by the coherent sound
radiation of the Cherenkov type.
\end{abstract}

\maketitle

\section{Introduction}

Primordial black holes could arise at the early stages of the
Universe evolution when the matter density was very high. But
which of them could survive since those times? The problem is that
too light primordial black holes have already evaporated due to
their thermal radiation. Well-known estimates \cite{Hawking}
demonstrate that the masses of the survivors should exceed
$10^{15}$ g.

As a step to study the possibility to detect the passage of so
small black hole through the Earth (or some other planet, or the
Moon), we analyze here the effects arising during such a passage
with a velocity comparable to that of the planet.

%In the present paper we discuss the possibility to detect such a
%light black hole by the effects created when it passes through the
%Earth (or some other planet, or the Moon) with velocity comparable
%to that of the planet. We show

It turns out that for a supersonic black hole, i.e. for that with
velocity exceeding the speed of sound, the main effect is a
coherent excitation of sound waves in the matter, the acoustic
analogue of Cherenkov radiation. These waves can be in principle
observed by seismic measurements.

\section{Coherent sound generation by a supersonic black hole}

 We start with the dynamics of mechanical deformations and excitation
 of sound waves caused by the passage of a primordial black hole through matter.
 The deformation can be conveniently described by the displacement
 vector $\bu$ with respect to the positions of the matter points in
 the absence of black hole. For small deformations $\bu$ satisfies
 the linear wave equation:
 \beq\label{motion}
 \ddot{\bu}-c_s^2 \Delta \bu=\bg.
 \eeq
Here ${\bf g}$ is the gravitational acceleration created by the
black hole, and $c_s$ is the speed of sound. We note that only the
longitudinal sound modes are excited by the gravitational force
since ${\bf g}$ is the gradient of a scalar potential: ${\bf
g}=\nabla \phi$.

 The gravitational acceleration {\bf g} satisfies the Gauss law:
 \beq
 \nabla\bg = - 4\pi G [M \delta (\br - \bv t) +\delta \rho].
 \eeq
Here $M$ and $\bv$ are the black hole mass and velocity,
respectively; the deviation $\de\rho$ of matter density from its
equilibrium value in the absence of a black hole is related to the
displacement vector as follows:
\[
\delta\rho = -\rho \nabla\bu\,.
\]
The divergence of Eq. (\ref{motion}) is
 \beq
 \ddot{\psi}-c_s^2 \Delta \psi - 4\pi G \rho\psi=-4\pi G M \delta (\br - \bv
 t)\,,
 \eeq
where $\psi=\nabla\bu$. The corresponding equation for the Fourier
transforms $\psi_{\bk}$ reads
 \beq\label{oscillation}
 \ddot{\psi}_{\bk}+\varepsilon \dot{\psi}_{\bk}+ c_s^2 k^2 \psi_{\bk}
 - 4\pi \rho\, G\psi_{\bk}=-4\pi  M G\exp(-i\bk\bv t)\,;
 \eeq
we have introduced here an infinitesimal damping $\varepsilon$
which corresponds to the retarded solution and is equivalent to a
small viscosity term in the wave equation. The solution of Eq.
(\ref{oscillation}) is
 \beq
 \psi_{\bk}=\frac{4\pi G M \exp(-i\omega t)}
 {\omega^2+i\omega\varepsilon-c_s^2 k^2+4\pi G \rho}\,,
 \eeq
 with $\omega=\bk\bv=k_{\parallel}v$.

The decelerating force applied to the black hole is equal to
 \beq
 M \bg=-4\pi i \rho\, M G\,\int \bk\,\psi_{\bk} \exp(i\omega t)
 \,\frac{d^3 \bk}{(2\pi)^3 k^2}\,.
 \eeq
The radiation intensity, or the rate of energy loss by the black
hole due to its elastic interaction with matter, is
 \[
 I_{el}=-M\,\bg\,\bv=\,(4\pi G M)^2 \rho\,i\,
 \int \frac{\omega}{\omega^2+i\omega\varepsilon-c_s^2 k^2+4\pi G \rho}\,
 \frac{d^3 \bk}{(2\pi)^3 k^2}\,
 \]
\beq
\quad \quad \quad \quad \quad \quad \quad  =\,(4\pi G M)^2
\rho\,i\,\int \frac{\omega}{\omega^2+i\omega\varepsilon-c_s^2
k^2+4\pi G \rho}\,
 \frac{d\omega\,d^2 \bk_{\perp}}{(2\pi)^3 v k^2}\,.
 \eeq
Now we use well-known relation
\[
\frac{1}{x+i\varepsilon}= P\frac{1}{x} - i\,\pi\,\delta(x)\,,
\]
where $P$ is the principal value of an integral. Here the
contribution of the principal value is an odd function of $\omega$
and therefore vanishes after integration over $\omega$. Thus, we
obtain
\beq\label{iel0}
I_{el} = 2(G M)^2 \rho\,
 \int \frac{d^2 \bk_{\perp}}{v k^2}\,\int_{-\infty}^\infty
 \delta(\omega^2(1-c_s^2/v^2)-c_s^2 k_\perp^2+4\pi G
 \rho)\,\omega\,d\omega\,.
 \eeq
In our case the black hole moves faster than sound, $v>c_s$, and
the $\delta$-function gives a non-vanishing contribution only if
$k_\perp>\omega_p/c_s$, where $\omega_p^2=4\pi\,G\rho$. Then,
\beq\label{iel}
 I_{el}=\frac{2\pi(G M)^2 \rho}{v\,(1-c_s^2/v^2)}\,
 \int_{(\omega_p/c_s)^2}^{k_1^2}\frac{dk_{\perp}^2}{k^2}\,=2\pi(G M)^2 \rho/v\,
 \int_{(\omega_p/c_s)^2}^{k_1^2}\frac{dk_{\perp}^2}{k_\perp^2-\omega^2_p/v^2}\,=
 2\pi(G M)^2 \rho/v\,
 \int_{(\omega_p/c_s)^2}^{k_1^2}\frac{dk_{\perp}^2}{k_\perp^2}\,;
 \eeq
we recall here that $v \gg c_s $. This integral diverges
logarithmically at large momentum transfers from the black hole to
the matter, when the scattering gets inelastic. Fortunately, if we
are interested in the total energy losses, elastic plus inelastic
(see below), the exact value $k_1^2$ of such critical momentum
transfers is inessential.

\section{Inelastic scattering. Total mechanical losses}

We start here with the opposite limiting case, that of large
momentum transfers, when the matter can be considered as a
collection of free particles (of mass $m$, number density $n$, and
mass density $\rho = m n$). The differential cross-section for
scattering of a black hole on such a particle is
\beq
\frac{d\si}{d\Om}\,=\,\frac{1}{4}\left(\frac{GM}{v^2}\right)^2
\frac{1}{\sin^4\theta/2}\,.
\eeq
The corresponding decelerating force looks formally as \beq F = \,
n m v^2 \,\frac{1}{4}\left(\frac{GM}{v^2}\right)^2 2\pi
\int^\pi_0\,\frac{d\theta \sin\theta
(1-\cos\theta)}{\sin^4\theta/2}\,= 2\pi
\rho\,\frac{(GM)^2}{v^2}\,\int^\pi_0\,d\theta\,
\frac{\cos\theta/2}{\sin\theta/2}\,.
\eeq
In fact, the typical scattering angles here are small, so that
\[
\int^\pi_0\,d\theta\,\frac{\cos\theta/2}{\sin\theta/2}\,
=\,\ln\frac{\theta_{max}^2}{\theta_{min}^2}\, =\,\ln\frac{k_{\perp
max}^2}{k_{\perp min}^2}\,.
\]
Obviously, $k_{\perp max} \simeq 1/a$, where $a$ is the typical
interatomic distance in the matter. As to the minimum momentum
transfer $k_{\perp min}$ at which one can neglect the binding of
the matter constituents, it coincides as obviously, at least in
the order of magnitude, with the maximum momentum transfer $k_1^2$
at which the interaction of a black hole with matter remains
elastic (see equation (\ref{iel})). Thus, the rate of inelastic
energy loss by a black hole is
\beq\label{in}
I_{inel} = F v = 2\pi(G M)^2 \rho/v\,
 \ln\frac{1}{k_1^2 a^2}
 \eeq

And finally, the total rate of energy loss is
\beq\label{tot}
I_{tot} = I_{el} + I_{inel} = 4\pi(G M)^2 \rho/v\,
\ln\frac{c_s}{\omega_p\, a}\,.
\eeq
With the accepted logarithmic accuracy, this total rate is
independent of the critical momentum transfer $k_1$. On the other
hand, for any reasonable choice of $k_1$, the elastic energy loss
dominates strongly, $I_{el} \gg I_{inel}$. It is worth mentioning
also that the logarithm in (\ref{tot}) is really large, about 35.
An expression for $I_{tot}$ close to (\ref{tot}), but with a
different logarithmic factor, was obtained previously by K.
Penanen \cite{Penanen}.

To estimate the energy $\De E$ released by a black hole passing
through the Earth, this rate should be multiplied by the time of
the passage, $\tau = L/v$. For numerical estimates we assume that
the equilibrium density of matter is $\rho = 6$~g/cm$^3$, the path
$L$ is about the Earth diameter, $L \sim 10^4$ km, and the
velocity of black hole is $v \sim 30$ km/s. At last, for a black
hole with mass $M  \sim 10^{15}$ g this energy loss constitutes
about
\beq\label{rel}
\De E \sim 4\times 10^{9}\; {\rm J}\,.
\eeq

Let us note that this energy is much smaller than that released at
the explosion of a 10 kiloton atomic bomb
\beq\label{bomb}
\De E_{bomb} \sim 5 \times 10^{13}\; {\rm J}\,.
\eeq
Besides, when comparing the energy released by a black hole (not
only (\ref{rel}), but also some other contributions to it
considered below) with the energy of an atomic bomb, one should
keep in mind that the source of $\De E_{bomb}$ is practically
point-like, while $\De E$ is spread along a path $L~\sim~10^4$~km.

\section{Conversion of black hole radiation into sound waves}

One more source of the energy transfer from a light black hole to
the matter (though not its kinetic energy discussed above, but the
internal one) is the black hole radiation. Of course, for our
purpose we have to consider the emission of $\ga$ and $e^\pm$ only
(but not gravitons and neutrinos). Using the results of
\cite{Page}, we obtain under the same assumptions ($M  \sim
10^{15}$ g, $L \sim 10^4$~km, and $v \sim 30$ km/s) the following
estimate for the total radiation loss of such black hole:
\beq\label{rad}
\De E_{rad} \sim 1.5 \times 10^{12}\; {\rm J}\,.
\eeq

One of the possible mechanisms for the conversion of radiation
into sound waves, which permits of rather reliable theoretical
analysis, is as follows. The radiation absorbed by matter
increases the temperature along the path of the source. This
results in the inhomogeneous and non-stationary thermal expansion
of the matter and thus in the emission of acoustic waves. The
matter is treated as a liquid (the case of a solid medium could be
considered analogously), and the well-known relations
\beq
\dot{\rho}+\divergence \rho \bv = 0, \;\;\; \rho \dot{\bv}=-\nabla
p
\eeq
result in the following equation
\beq \label{wave}
\ddot{\rho}-\Delta p = 0, \eeq for density $\rho$ and pressure
$p$. The variations of density, pressure, and temperature are
related as follows:
\beq
\delta \rho=\left(\frac{\partial \rho}{\partial p}\right)_T \delta
p +\left(\frac{\partial \rho}{\partial T}\right)_p \delta T
=\frac{1}{c^2_s}\delta p-\rho \beta \delta T ;
\eeq
here $\beta= - 1/\rho\; (\partial \rho/\partial T)_p$ is the
coefficient of thermal expansion. This allows one to eliminate
$\rho$ from (\ref{wave}):
\beq
\frac{1}{c_s^2}\ddot{p}-\Delta p -\rho \beta \ddot{T} = 0.
\eeq
Neglecting the thermal conductivity, we rewrite $\rho \dot T$ as
$W/C$, where $W$ is the power density and $C$ is the specific
heat. Thus, such effect due to heating is described by the
following equation for pressure (derived previously in Ref.
\cite{wes}):
\beq \label{source}
\frac{1}{c_s^2}\ddot{p}-\Delta p = \frac{\beta}{C} \dot{W}.
\eeq
A black hole can be treated as a point-like source of radiation
with intensity $I$, so that in our case $W=I \delta (\br - \bv
t)$.

Let us consider now the mechanical energy of the matter:
\beq
E_m=\int \left(\frac{p^2}{2\rho c_s^2} + \frac{\rho v^2}{2}\right)
dV.
\eeq
The intensity of the sound radiation coincides with the rate of
energy loss:
\beq
\frac{dE_m}{dt}=\int \left(\frac{p \dot{p}}{\rho c_s^2} + \rho \bv
\dot{\bv}\right) dV = \int \left(\frac{p \dot{p}}{\rho c_s^2} -
\bv \nabla p\right) dV.
\eeq
We use here the Euler equation $\rho \dot{\bv}=-\nabla p$ to
rewrite the last term. Then, integrating by parts and using the
continuity equation we arrive at:
\beq
\frac{dE_m}{dt}= \int \frac{p}{\rho} \left(\frac{\dot{p}}{c_s^2} -
\dot{\rho}\right) dV = \int \frac{\beta p W}{\rho C} dV.
\eeq
With a point-like source $W~=~I~\delta(\br - \bv t)$, the
integration is performed easily:
\beq
\frac{dE_m}{dt}= \frac{\beta I}{\rho C}\; p\,(\bv t),
\eeq
where p(\bv t) is the pressure at the point of the black hole
location. Now the Fourier transformed Eq. (\ref{source}) is
\beq
p_{\omega,\bk}=\frac{\beta I}{C} (-i\omega) \frac{2\pi
\delta(\omega -\bk\bv)}{k^2 -(\omega+i\varepsilon)^2/c_s^2}.
\eeq
Here we have introduced again an infinitesimal damping
$\varepsilon$. With $\varepsilon \rightarrow 0$, the frequency
Fourier component of the pressure at the black hole location is
\[
p_\omega = \int \frac{d\bk}{(2\pi)^3} p_{\omega,\bk} =\frac{\beta
I}{C} \int \frac{d\bk}{(2\pi)^2}\frac{-i\omega\delta(\omega
-\bk\bv)}{k^2-(\omega+i\varepsilon)^2/c_s^2} \longrightarrow
\frac{\beta I \omega}{C} \int
\frac{d\bk}{(2\pi)^2}\,\pi\,\delta(\omega
-\bk\bv)\,\delta\left(k^2-\omega ^2/c_s^2\right)
\]
\beq\label{pom}
= \frac{\beta I \omega}{4 v C} \int_0^\infty
dk_\perp^2\delta\left(k_\perp^2 + \omega^2/v^2
-\omega^2/c_s^2\right) = \frac{\beta I \omega}{4 v C}\,.
\eeq

The intensity of the sound radiation at given frequency $\omega$
is related to $p_\omega$ as follows:
\beq
\frac{dE_m}{dt d\omega}\,d\omega = 2\,\frac{\beta I}{\rho C}\,
p_\omega \,\frac{d\omega}{2\pi}\,;
\eeq
the overall factor 2 in this expression corresponds to the fact
that frequencies of both signs, $\omega$ and $-\omega$, are taken
here into account. Finally, we arrive at the following result for
the spectral intensity of the sound waves:
\beq
\frac{dE_m}{dt d\omega}=\left(\frac{\beta I}{C}\right)^2
\frac{\omega}{4\pi v \rho}.
\eeq

The total energy radiated in this way at the frequency $\omega$,
during the passage of a black hole through the Earth, can be
conveniently written as
\beq\label{Em}
\frac{dE_m}{d\omega}=\left(\frac{\beta \Delta E_{rad}}{C}\right)^2
\frac{\omega}{4\pi L \rho}\,;
\eeq
we go over in this expression from the intensity $I$ of the black
hole radiation to the total energy $\Delta E_{rad}$ emitted during
the passage through the Earth: $\Delta E_{rad}=I\,L/v$.

Curiously, this sound radiation occurs only if $v > c_s$, as it is
obvious from the last line of Eq. (\ref{pom}). In other words,
this is also a sort of Cherenkov effect.

\section{Matter excitations and sensitivity of seismic detectors}

To detect a mini-black hole passing through the Earth, one has to
study seismic vibrations induced by this passage. The sensitivity
of appropriate seismic detectors is confined to the frequencies in
the interval around $\om_{min} \sim 1$ Hz and $\om_{max} \sim 100$
Hz.

To determine the frequency distribution of the acoustic Cherenkov
radiation, we come back to formula (\ref{iel0}) and perform at
first integration over $\bk_{\perp}$. With $v \gg c_s$ and $\omega
\gg \omega_p$, the result is
\beq
dI_{el}\,=\,4\pi(GM)^2(\rho/v)\,\frac{d\om}{\om}\,.
\eeq
Thus, the energy of the vibrations excited in the frequency
interval $\om_{min} \div \,\om_{max}$ is
\beq
\De E^{\om} = 4\pi(G M)^2 L\rho/v^2\,
\ln\frac{\om_{max}}{\om_{min}}\,.
\eeq
For the discussed frequency interval, 1 $\div$ 100 Hz, it
constitutes numerically
\beq\label{relom}
\De E^{\om} \sim \,5 \times 10^{8}\; {\rm J}\,,
\eeq
or about 1/10 of the total energy (\ref{rel}).

As to the seismic waves generated by the black hole radiation,
their total energy for frequencies $\omega < \omega_{max}$ is,
according to (\ref{Em}),
\beq
E^{\om}_m =\left(\frac{\beta \Delta E_{rad}}{C}\right)^2
\frac{\omega_{max}^2}{8\pi L \rho}.
\eeq
With $C=1\,{\rm J}\, {\rm g}^{-1}\,{\rm K}^{-1}$ and $\beta=0.5
\cdot 10^{-4}\, {\rm K}^{-1}$, we obtain $E^{\om}_m \sim 40\,{\rm
J}$. So, this effect is much less than that of the Cherenkov sound
radiation (\ref{relom}).

There is in fact one more mechanism by which the energy of black
hole radiation is transformed into the mechanical energy of
matter. It is as follows. Of course, in the rest frame of a black
hole its radiation is isotropic, so that the total momentum of
radiated particles is equal to zero. However, it is not so in the
rest frame of the Earth. In it the momentum carried away by
particles radiated by a black hole with velocity $v$ per unit time
can be estimated as $I v/c^2$ (here $c$ is the speed of light).
All this momentum is absorbed by the matter together with the
radiation itself. Therefore, the effective force of interaction
between the black hole and matter is
\beq
F \sim \frac{I v}{c^2} \sim \frac{\Delta E_{rad}\, v^2}{c^2 L}\,.
\eeq
Thus, the total energy transferred due to this pressure from the
black hole to matter along the path $L$ can be estimated as
\beq\label{epr}
\Delta E_{pr} \sim F\,L \sim  \Delta
E_{rad}\,\frac{v^2}{c^2}\,\sim 10^{-8}\,\Delta E_{rad}\,\sim
10^4\,{\rm J}\,.
\eeq

However, the portion of energy released in this way in the
frequency region $\omega < \omega_{max}$ $\sim$ 100 Hz is much
less. Its crude estimate looks as follows. The typical absorption
length $r_0$ for $\ga$ and $e^\pm$ is in our case about 3 cm.
Simple estimates demonstrate that the relative concentration of
the defects in matter created by this radiation is small
everywhere. (In this respect as well, our problem of the seismic
waves, generated by the black hole radiation, differs from that
for the underground explosion of an atomic bomb: in the last case
the region of complete destruction of the matter is measured at
least by meters.) Thus, in the present case $r_0$ is the only
length scale at our disposal. This region of radius $r_0$
propagates in the matter with the velocity $v$ of a black hole.
Then the typical frequency of thus created perturbations in the
matter can be estimated as
\beq
\om_0 \sim v/r_0 \sim 10^6\; {\rm Hz}.
\eeq
The frequencies $\om$, we are interested in, are much smaller,
$\om_{max} \ll \om_0$. It looks natural to assume that in our low
frequency (``infrared'') region the radiation intensity is
governed by the phase space considerations. With the essentially
two-dimensional propagation of the perturbation around the line of
flight of a black hole, the phase space can be considered also as
a two-dimensional one. Then one can assume that the frequency
spectrum in the low-frequency region of interest looks as
$\om^2/\om_0^2$, and the ``useful'' portion of black hole
radiation, which directly induces seismic vibrations in the
frequency interval 1 $\div$ 100 Hz, can be estimated as
\beq
\De E_{pr}^{\om} \sim
E_{rad}\,\frac{v^2}{c^2}\,\frac{\om_{\max}^2}{\om_0^2}\,,
\eeq
which is negligibly small. Anyway, even the total energy
(\ref{epr}) transferred in this way is less than the useful part
(\ref{relom}) of the sound Cherenkov radiation.

One more effect we wish to mention is as follows. It was pointed
out long ago \cite{ask} that, due to the positron annihilation and
the capture of Compton- and $\delta$-electrons by a cosmic-ray
electron-photon shower, an excess of electrons arises in this
shower. Due to the excess, a shower gets effectively charged,
creates correspondingly a macroscopic electromagnetic field, and
radiates coherently.

Obviously, analogous effects take place in principle during the
passage of a radiating black hole through the Earth (of course,
the intensities of electrons and positrons emitted by a black hole
itself are equal). Let us estimate the field created in this way
by a black hole. We assume that the excess of electrons created
here per second is about the same as the total production rate of
particles by a black hole $\dot{N}$. Then the magnetic field at
the distance on the order of the absorption length $r_0$ from the
black hole can be estimated as
\beq
B \sim e\frac{v}{c}\,\frac{\dot{N}}{c r_0}\,.
\eeq
It falls down as $1/r^2$ with the distance from the black hole.
With $\dot{N}~\sim~10^{21}$~s$^{-1}$, $v \sim 30$ km/s, and $r_0
\sim 3$ cm, this magnetic field is about 0.3 Gs only. The
possibility to use such a signal for detecting the passage of a
light black hole through the Earth does not look realistic.

To summarize, the seismic signal of the passage of a light black
hole through the Earth in the frequency interval 1 $\div$ 100 Hz
is strongly dominated by the sound Cherenkov radiation, and its
total energy in this interval can be estimated as
\beq\label{relom1}
\De E^{\om} \sim \,5 \times 10^{8}\; {\rm J}\,.
\eeq

We wish to point out in conclusion that though the effects of
radiation damage contribute negligibly into the seismic signal,
they can create quite a distinct pattern in crystalline material.
The dose deposited is estimated as
\beq
\frac{\De E_{\rm rad}}{\rho L r_0^2}\; \sim \,10^{5}\; {\rm Gy}
\quad \quad (\, 1 {\rm Gy}\; ({\rm Gray}) = 1\;{\rm J/kg}\,)\,.
\eeq
It creates a long tube of heavily radiative damaged material,
which should stay recognizable for geological time.

\end{document}